\documentstyle[11pt,axodraw]{article}

    \advance\textheight by \topskip

\begin{document}
\hfill
UM-TH-97-01 

\hfill
NIHHEF-97-001
\begin{center}
{\huge \bf The four-loop $\beta$-function in 
Quantum Chromodynamics} \\[8mm] 
 T. van Ritbergen$^a$, J.A.M. Vermaseren$^b$, S.A. Larin$^{c}$ \\ [3mm]
\begin{itemize}
\item[$^a$]
 Randall Laboratory of Physics, University of Michigan,\\
 Ann Arbor, MI 48109, USA
\item[$^b$]
 NIKHEF-H, P.O. Box 41882, \\ 1009 DB, Amsterdam, The Netherlands \\
\item[$^c$]
 Institute for Nuclear Research of the
 Russian Academy of Sciences,   \\
 60th October Anniversary Prospect 7a,
 Moscow 117312, Russia
\end{itemize}
\end{center}

\begin{abstract}
We present the analytical calculation of the four-loop QCD $\beta$-function
 within the minimal subtraction scheme.
\end{abstract}

\newpage
The renormalization group $\beta$-function in
Quantum Chromodynamics (QCD) has a history of more than 20 years. 
The calculation of the one-loop $\beta$-function in
QCD has lead to the discovery of asymptotic freedom
in this model and to the establishment of QCD as the theory 
of strong interactions \cite{gvp}.

The two-loop QCD $\beta$-function was calculated in \cite{2l}.
Calculations of the three-loop QCD $\beta$-function 
were done in \cite{tvz,lv} within the minimal subtraction (MS)
scheme \cite{h}. The MS-scheme belongs to the class of massless schemes
where the $\beta$-function does not depend on masses of the theory
and the first two coefficients of the $\beta$-function are
scheme-independent.

In this article we present the analytical four-loop
result for the QCD $\beta$-function. 
Throughout the calculations we use dimensional regularization \cite{hv}
and the MS-scheme. 
The dimension of 
space-time is defined as $D=4-2\varepsilon$, 
where $\varepsilon$ is the regularization parameter
fixing the deviation of the space-time
dimension from its physical value 4.

The Lagrangian for a massless non-abelian Yang-Mills theory
with fermions is
\begin{eqnarray}
\label{lagrangian}
 L & = & -\frac{1}{4} G^{a}_{\mu\nu} G^{a\mu\nu} +
 i \sum_q \overline{\psi_q} D\!\!\!\!\slash\,  \psi_q +L_{gf}+L_{gc}
 \nonumber \\
 G_{\mu\nu}^{a}  & = & \partial_\mu A_\nu^{a} -\partial_\nu A_\mu^a
      + g f^{abc} A_\mu^b A_\nu^c  \nonumber \\
 \left[ D_{\mu}\right]_{ij} & = &
         \delta_{ij} \partial_\mu -i g A_\mu^a [T^a]_{ij}
\end{eqnarray}
where the gauge-fixing and gauge-compensating parts of the Lagrangian
in the covariant gauge are
\begin{eqnarray}
L_{gf} & = & -\frac{1}{2\xi}(\partial^\mu A_\mu^a)^2 \nonumber \\ 
L_{gc} & = & \partial^{\mu}\overline{\omega}^a(\partial_\mu\omega^a
                 -g f^{abc} \omega^b A_\mu^c)
\end{eqnarray}
The fermion fields (the quark fields in QCD) $\psi_q$ transform as 
the fundamental 
representation of
 a compact semi-simple Lie group,  $q=1,..., n_f$ is the flavour index.  
The Yang-Mills fields
(gluons in QCD) $A^a_\mu$ 
transform as the adjoint representation of this group.
$\omega^a$ are the ghost fields, and $\xi$ is the gauge parameter
of the covariant gauge.

$T^{a}$ are the generators of the fundamental representation
and  $f^{abc}$ are the structure constants of the Lie algebra,
\begin{equation}
T^aT^b-T^bT^a = i f^{a b c} T^c
\end{equation}
In the case of QCD we have the Lie group $SU(3)$ but we will perform 
the calculation for an arbitrary compact semi-simple Lie group $G$. Since 
the $\beta$-function does not depend on masses
in the MS-scheme, we will consider the massless theory.  

The definition of the 4-dimensional $\beta$-function is:
\begin{eqnarray}
\label{rengroup}
\frac{\partial a_s }{\partial \, \ln \mu^2}  & = &
\beta(a_s) \nonumber \\
& = & -\beta_0 a_s^2 - \beta_1 a_s^3
-\beta_2 a_s^4
-\beta_3 a_s^5 + O(a_s^6) \hspace{1cm}
\end{eqnarray}
in which $a_s=\alpha_s/4\pi=g^2/16\pi^2$, $g=g(\mu^2)$ 
is the renormalized strong
coupling constant of the standard QCD Lagrangian of eq.(\ref{lagrangian}).
$\mu$ is the 't Hooft
unit of mass, the renormalization point in the MS-scheme.

To calculate the $\beta$-function we need to calculate 
the renormalization constant $Z_{a_s}$ of the coupling constant
\[ a_{B} = Z_{a_s} a_s\]
where $a_{B}$ is the bare (unrenormalized) charge.
We obtain this renormalization constant in the 4-loop order by 
calculating the following three renormalization constants of the Lagrangian:
$Z_{hhg}$ for the ghost-ghost-gluon vertex,
$Z_h$ for the inverted ghost propagator and $Z_g$ for the inverted gluon
propagator. Then $Z_{a_s}=Z_{hhg}^2/Z_h^2/Z_g$.
This is from a calculational point of view one of the simplest 
(and most straightforward) ways to obtain this renormalization constant 
at higher orders. However, several other choices are possible such as 
calculating the renormalization factors of the
quark propagator, the gluon propagator and the quark-gluon vertex.
One could also use the background field method \cite{abbott}, 
which reduces the calculation of the $\beta$-function
to the calculation of the gluon propagator only. However, we should note
that the Feynman rules are more complicated in that case, 
and this complication would lead
at the 4-loop level to a complexity of the calculations that is comparable
to our more standard approach. 

The expression of the $\beta$-function via $Z_{a_s}$ is given 
by the following chain of equations

\[ \frac{ d ( a_{B} \mu^{2\varepsilon})}{d \ln\mu^2} = 0 
   = \varepsilon Z_{a_s} a_s \mu^{2\varepsilon} +
        \frac{\partial Z_{a_s} }{\partial a }
         \frac{d a_s}{d \ln \mu^2 } a_s \mu^{2\varepsilon}
        + Z_{a_s} \frac{d a_s}{d \ln \mu^2 } \mu^{2\varepsilon} \]
\begin{equation} \Rightarrow
 \frac{d a_s}{d \ln \mu^2 } = - \frac{ \varepsilon Z_{a_s} a_s}{
      \frac{\partial Z_{a_s} }{\partial a_s } a_s + Z_{a_s} }  =
-\varepsilon a_s-a_s\frac{\partial}{\partial a_s}\left(a_s Z_{a_s}^{(1)}\right)=
  -\varepsilon a_s + \beta (a_s) \end{equation}
where one uses the fact that the dimensional object
 $a_{B} \mu^{2\varepsilon}$ is invariant under the renormalization group
transformations.
$[-\varepsilon a_s + \beta (a_s)]$ is the $D$-dimensional
 $\beta$-function, $Z_{a_s}^{(1)}$ is the coefficient of the first 
$\varepsilon$-pole in $Z_{a_s}$ defined below.

Renormalization constants within the MS-scheme do not
depend on dimensional parameters (masses, momenta) \cite{collins} and
have the following structure:
\begin{eqnarray}
\label{zet}
Z_{a_s}(a_s)=1+\sum_{n=1}^{\infty}\frac{Z_{a_s}^{(n)}(a_s)}{\varepsilon^n},
\end{eqnarray}
Since $Z_{a_s}$ does not depend explicitly on $\mu$, the $\beta$-function
is the same in all MS-like schemes, i.e. within the class of 
renormalization schemes which differ by the shift of the parameter $\mu$.
That is why the $\beta$-function is the same in the MS-scheme \cite{h}
and in the $\overline{\rm MS}$-scheme \cite{bbdm}.

In general, the most straightforward way to obtain renormalization
constants is by multiplicative renormalization of the relevant Green functions
\begin{eqnarray}
\label{multren}
\Gamma_{\rm Renormalized}(a_s)=Z(a_s)~\Gamma_{\rm Bare}(a_{B}),
\end{eqnarray}
This direct approach was used for the independent calculation of the 
3-loop $\beta$-function in \cite{lv}. However the extension of this approach 
to 4-loops requires the calculation of 4-loop massless propagator type 
integrals which are still difficult to evaluate (at this moment).  

Another approach to find the renormalization constants $Z_i$ is to obtain them
as the sum of the counterterms of individual diagrams. 
This means that one applies the R-operation to each
individual diagram that contributes to a propagator or vertex function.
This allows a greater freedom in choosing the type of integrals that one 
needs to evaluate.
This approach (based on massless propagator type integrals)
was used for the first 3-loop calculation of the $\beta$-function \cite{tvz}.

For the calculation presented in this article we use massive integrals,
instead of the massless integrals used in previous calculations.
The calculation of renormalization constants within the MS-scheme 
can be reduced to the calculation of massive vacuum bubble integrals
(i.e. massive integrals with no external momenta)
using the general method of infrared rearrangement \cite{vladimirov}.
This method uses the property that within dimensional regularization
overall ultra-violet divergences are polynomial in external momenta
and masses, also for individual diagrams.
For the renormalization constants of the ghost-ghost-gluon vertex, 
ghost propagator and gluon propagator this means, 
that we can safely apply Taylor expansions
in the external ghost- and gluon momenta if we introduce 
a non-zero auxiliary mass $M$ for all internal propagators 
(also for the gluons).
It is understood that this auxiliary mass serves only as an infrared cutoff 
parameter that is nullified {\em after} renormalization of 
the individual diagrams. For simplicity, we introduce the mass only
in the denominators of the propagators, not in the numerators.
The difference between the overall divergences of the diagrams with and 
without the mass is a term that is polynomial in this mass and vanishes
when $M$ is nullified.

The procedure of renormalization with an auxiliary mass works well 
for individual diagrams. However the introduction of the mass $M$ 
in the gluon propagators spoils multiplicative renormalizability of the 
(massive) Green functions. 

Here we use an intermediate approach to renormalization in order
to get the 4-loop counterterms for the {\em sum} of the diagrams.
We compute poles in $\varepsilon$ of the corresponding 4-loop
massive diagrams. But we do not renormalize each diagram separately. 
The subtraction of subdivergences is done
for the whole sum of the 4-loop diagrams. This is done by means of adding  
to the sum of the 4-loop diagrams
the sum of the necessary bare diagrams of 1,2 and 3-loops
with all vertices replaced by effective vertices 
and all propagators replaced by effective propagators.
The effective vertices contain the necessary vertex renormalization 
constants up to the appropriate order in $a_s$ and similarly, the 
effective propagators contain the necessary propagator counterterms.
The various vertex and propagator renormalization constants 
that are needed in the effective vertices and propagators are 
already known from the lower order massless calculations 
(we emphasize that they are mass independent) except for the
overall uv divergence proportional to $M^2$ of the gluon
propagator, which is only needed up to 3-loops. 
 
Special routines for the symbolic manipulations program FORM \cite{form} 
were constructed to efficiently evaluate the 4-loop massive bubble integrals 
up to pole parts in $\varepsilon$
and correspondingly of the 3-loop massive bubbles up to finite parts.
For the 4-loop integrals, we only needed to deal with two master bubble 
topologies, see fig. 1.  
The various vertex and propagator diagrams were generated by means of 
the diagram generator QGRAF \cite{qgraph}. For the present calculation 
we evaluated of the order of 50.000 4-loop diagrams.

\begin{center} \begin{picture}(370,60)(0,10)

\SetWidth{1}
\SetScale{.56}

\CArc(50,50)(40,0,360)

\SetOffset(80,0)
\CArc(50,50)(40,0,360)
\Line(50,10)(50,90)

\SetOffset(160,0)
\CArc(50,50)(40,0,360)
\Line(50,50)(50,90)
\Line(50,50)(84.64101616,30)
\Line(50,50)(15.35898384,30)

\SetOffset(240,0)
\CArc(50,50)(40,0,360)
\CArc(50,50)(15,0,360)
\Line(50,65)(50,90)
\Line(62.99038106,42.5)(84.64101616,30)
\Line(37.00961894,42.5)(15.35898384,30)

\SetOffset(320,0)

\CArc(50,50)(40,0,360)
\Line(30,84.64101616)(70,15.35898384)
\Line(10,50)(43,50)
\Line(57,50)(90,50)
\Line(30,15.35898384)(46.5,43.93782217)
\Line(70,84.64101616)(53.5,56.06217783)

\end{picture} \\
\end{center}
\[\parbox{13cm}{ {\bf Figure 1.}  
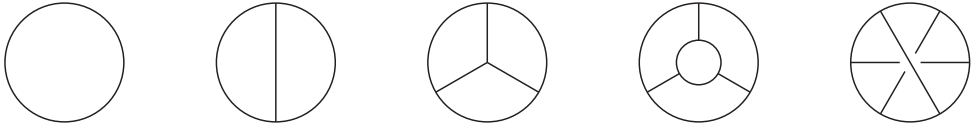
 The basic (master) vacuum bubble topologies up to 4 loops. 
 It is understood that all lines have the same non-zero mass, $M$. 
 There are two basic 4-loop topologies (the last one shown is non-planar).} \]

We obtained in this way the following result for the 
4-loop beta function in the $\overline{\rm MS}$-scheme
\renewcommand{\arraystretch}{1.3}
\begin{eqnarray} 
\label{eq:beta3}
\beta_{0} & = &  \frac{11}{3} C_A - \frac{4}{3} T_F n_f 
 \nonumber \\
 \beta_{1} & = &
 \frac{34}{3}C_A^2 - 4 C_F T_F n_f -\frac{20}{3} C_A T_F n_f
   \nonumber \\
 \beta_{2} & = &  \frac{2857}{54} C_A^3
 +2 C_F^2 T_F n_f - \frac{205}{9} C_F C_A T_F n_f  \nonumber \\
 & & - \frac{1415}{27} C_A^2 T_F n_f
 + \frac{44}{9} C_F T_F^2 n_f^2
  + \frac{158}{27} C_A T_F^2 n_f^2  
   \nonumber \\
\beta_{3} & = &
    C_A^4 \left( \frac{150653}{486} - \frac{44}{9} \zeta_3 \right)   
    +  C_A^3 T_F n_f  
      \left(  - \frac{39143}{81} + \frac{136}{3} \zeta_3 \right)
\nonumber \\ & &
    + C_A^2 C_F T_F n_f 
\left( \frac{7073}{243} - \frac{656}{9} \zeta_3 \right)
   +     C_A C_F^2 T_F n_f 
      \left(  - \frac{4204}{27} + \frac{352}{9} \zeta_3 \right)
\nonumber \\ & &
   + 46 C_F^3 T_F n_f 
   +  C_A^2 T_F^2 n_f^2 
      \left( \frac{7930}{81} + \frac{224}{9} \zeta_3 \right)
    +  C_F^2 T_F^2 n_f^2 
      \left( \frac{1352}{27} - \frac{704}{9} \zeta_3 \right)
\nonumber \\ & &
    +  C_A C_F T_F^2 n_f^2 
      \left( \frac{17152}{243} + \frac{448}{9} \zeta_3 \right)
  + \frac{424}{243} C_A T_F^3 n_f^3 
   + \frac{1232}{243} C_F T_F^3 n_f^3  
\nonumber \\ & &
       +  \frac{d_A^{a b c d}d_A^{a b c d}}{N_A }  
             \left(  - \frac{80}{9} + \frac{704}{3} \zeta_3 \right)
       + n_f \frac{d_F^{a b c d}d_A^{a b c d}}{N_A }  
            \left(   \frac{512}{9} - \frac{1664}{3} \zeta_3 \right)
\nonumber \\ & &
       + n_f^2 \frac{d_F^{a b c d}d_F^{a b c d}}{N_A }   
            \left(  - \frac{704}{9} + \frac{512}{3} \zeta_3 \right)
 \label{mainbeta} \end{eqnarray}  \\
Here $\zeta$ is the Riemann zeta-function ($\zeta_3 = 1.202056903\cdots $).
 $[T^a T^a]_{ij} = C_F \delta_{ij}$ and
 $f^{a c d} f^{b c d} = C_A \delta^{ab}$ are the Casimir operators of the
fundamental and the adjoint representation of the Lie algebra.
tr$(T^a T^b) = T_F \delta^{a b}$ is the trace normalization of the
fundamental representation. $N_A$ is the number of generators of the group
(i.e. the number of gluons) and $n_f$ is the number of
 quark flavours. We expressed the higher order 
group invariants in terms of contractions between the following fully 
symmetrical tensors:
\begin{eqnarray}
 d_F^{a b c d} & = & \frac{1}{6 } {\rm Tr }  \left[
   T^a T^b T^c T^d
 + T^a T^b T^d T^c 
 + T^a T^c T^b T^d  \right. \nonumber \\
 & & \left. \hspace{4mm}
 + T^a T^c T^d T^b 
 + T^a T^d T^b T^c 
 + T^a T^d T^c T^b  \hspace{1mm}
 \right]\\
 d_A^{a b c d} & = & \frac{1}{6} {\rm Tr }  \left[
   C^a C^b C^c C^d
 + C^a C^b C^d C^c 
 + C^a C^c C^b C^d  \right. \nonumber \\
 & & \left. \hspace{4mm}
 + C^a C^c C^d C^b 
 + C^a C^d C^b C^c 
 + C^a C^d C^c C^b  \hspace{1mm}
 \right]
\end{eqnarray}
where the matrices $[C^a]_{bc} \equiv - i f^{abc}$ are the generators in the
adjoint representation. The result of eq. (\ref{mainbeta}) 
is valid for an arbitrary semi-simple compact Lie group. 
The result for QED (i.e. the group U(1)) 
is included in eq.(\ref{mainbeta}) by substituting
$C_A = 0$, $d_A^{a b c d} = 0$, $C_F = 1$, $T_F = 1$, $d_F^{a b c d}=1$,
$N_A = 1$.
This result for QED agrees with the literature \cite{qed}. 
A second independent check
of eq. (\ref{mainbeta}) is provided by the calculation
\cite{gracey} where the large-$n_f$ terms for 
the QCD beta-function were calculated
in all orders of the coupling constant. Our $n_f^3$ terms agree 
with \cite{gracey}.

The result of eq.(\ref{mainbeta}) is obtained in an arbitrary
covariant gauge for the gluon field.
This means that we keep
the gauge parameter $\xi$ that appears in the gluon propagator
$ i$  $ [-g^{\mu\nu}+(1-\xi)
 q^{\mu}q^{\nu}/(q^2+i\epsilon)]/(q^2+i\epsilon)$  as a
free parameter in the calculations. The explicit cancellation of the gauge
dependence in the $\beta$-function gives an important check of the results.
The results for individual
diagrams that contribute to the $\beta$ function also contain (apart
from the constant $\zeta_3$) the constants $\zeta_4$, $\zeta_5$
and several other constants specific for massive vacuum integrals.
The cancellation of these constants at various stages in the calculation
provides additional checks of the result.

For the standard normalization of the SU($N$) generators we find
the following expressions for the colour factors

\[ T_F = \frac{1}{2}, \hspace{.5cm} C_A = N, \hspace{.5cm}
 C_F = \frac{N^2-1}{2 N} ,
 \hspace{.5cm}  \frac{d_{A}^{a b c d} d_{A}^{a b c d}}{N_A}
                         = \frac{N^2(N^2+36)}{24}, \]
\[  \frac{d_F^{a b c d}d_A^{a b c d}}{N_A}  =
                          \frac{ N(N^2+6)}{48}, \hspace{.5cm}
  \frac{d_F^{a b c d}d_F^{a b c d}}{N_A}  =
                  \frac{N^4-6N^2+18}{96 N^2} \]
Substitution of these colour factors for $N=3$ into eq.(\ref{mainbeta}) yields
the following result for QCD

\begin{eqnarray}
\renewcommand{\arraystretch}{ 1.3}
 \beta_0 & = & 11 - \frac{2}{3} n_f   \nonumber\\
\beta_1 & = & 102 - \frac{38}{3} n_f  \nonumber \\
 \beta_2 & = & \frac{2857}{2} - \frac{5033}{18} n_f + \frac{325}{54}
  n_f^2 \nonumber \\
 \beta_3 & = &  \left( \frac{149753}{6} + 3564 \zeta_3 \right)
        - \left( \frac{1078361}{162} + \frac{6508}{27} \zeta_3 \right) n_f
  \nonumber \\ & &
       + \left( \frac{50065}{162} + \frac{6472}{81} \zeta_3 \right) n_f^2
       +  \frac{1093}{729}  n_f^3
\end{eqnarray}
Or in a numerical form
\begin{eqnarray}
 \beta_0 & \approx & 11 - 0.66667 n_f \nonumber \\
 \beta_1 & \approx & 102 - 12.6667  n_f \nonumber \\
 \beta_2 & \approx & 1428.50 - 279.611 n_f +6.01852 n_f^2 \nonumber \\
 \beta_3 & \approx & 29243.0  - 6946.30 n_f  +405.089 n_f^2  +1.49931 n_f^3
\end{eqnarray}
We note that $\beta_3$ is positive for all positive values of $n_f$. 

It is interesting to compare our result with a recent prediction 
\cite{eks} for the 4-loop coefficient of the QCD $\beta$-function
in the $\overline{\rm MS}$-scheme using Pad\'e Approximants. 
For $n_f=3$ this prediction is within a factor 2 of the exact result
(for $n_f=5$ this factor is about 9).

Considering that one might want to use the results of 
equation~\ref{eq:beta3} for different groups or representations we will 
also express its constants in a different way~\cite{patera}. In general one 
can write for a representation $R$ of a simple Lie group
\begin{eqnarray}
  d_R^{abcd} & = & I_4(R)d^{abcd} + \frac{I_{2,2}(R)}{3}(
\delta^{ab}\delta^{cd}+\delta^{ac}\delta^{bd}+\delta^{ad}\delta^{bc}) 
    \nonumber
\end{eqnarray}
in which the tensor $d$ is now traceless. The only exceptions to this are the 
spinor representations of SO(8) for which there are two fully symmetric 
traceless tensors with 4 indices. We will not consider this special case 
here. The normalization of the tensor $d$ is fixed by the definition of 
$I_4(F)$. Contraction with $\delta^{ab}\delta^{cd}$ gives
\begin{eqnarray}
  I_{2,2}(R) & = & \frac{3}{N_A+2}J_{2,2}(R) \nonumber \\
  J_{2,2}(R) & = & (\frac{N_A}{N_R}-\frac{1}{6}
			\frac{I_2(A)}{I_2(R)})(I_2(R))^2 \nonumber \\
             & = & T_R ( C_R-\frac{1}{6}C_A)
\end{eqnarray}
with $I_2(R) = T_R$, $I_2(A) = T_A = C_A$ and $N_A T_R = C_RN_R$. We use here 
that $N_R$ and $N_A$ are the dimensions of the representation $R$ and the 
adjoint representation respectively. For the adjoint representation we have 
that $J_{2,2}(A) = 5C_A/6$.

\noindent After this the following identities hold:
\begin{eqnarray}
\frac{d_R^{a b c d}d_R^{a b c d}}{N_A } & = &
		 (I_4(R))^2\frac{d^{a b c d}d^{a b c d}}{N_A }
		+\frac{3}{N_A+2}(J_{2,2}(R))^2 \nonumber \\
\frac{d_R^{a b c d}d_A^{a b c d}}{N_A } & = &  
		 I_4(R)I_4(A)\frac{d^{a b c d}d^{a b c d}}{N_A }
		+\frac{3}{N_A+2}J_{2,2}(R)J_{2,2}(A) \\
\frac{d_A^{a b c d}d_A^{a b c d}}{N_A } & = &  
		 (I_4(A))^2\frac{d^{a b c d}d^{a b c d}}{N_A }
		+\frac{3}{N_A+2}(J_{2,2}(A))^2 \nonumber
\end{eqnarray}
These $d$'s have some nice properties. They are zero for all exceptional 
groups and for SU(3). In addition they are representation independent. 
Hence we need to give them only for the classical groups:
\begin{eqnarray}
	d^{abcd}d^{abcd}(SU(N)) & = & \frac{N_A(N_A-3)(N_A-8)}{96(N_A+2)}
					\nonumber \\
	d^{abcd}d^{abcd}(SO(N)) & = & \frac{N_A(N_A-1)(N_A-3)}{12(N_A+2)}
					\\
	d^{abcd}d^{abcd}(SP(N)) & = & \frac{N_A(N_A-1)(N_A-3)}{192(N_A+2)}
					\nonumber
\end{eqnarray}
For the overall normalization we define the constant $b$ with the relation 
\begin{equation}
Tr\left[C^\alpha C^\beta\right] = b\ g\ \delta^{\alpha\beta}
\end{equation}
in which $g$ is the dual Coxeter number. 
The factor b is related to the normalization factor $a$ in the 
article by Cvitanovic~\cite{cvitanovic}. The relation is $b=2a$ for the 
groups SU($N$) and SP($N$) and $b=a$ for SO($N$). 
The canonical choice of $b$ is 1 for all groups. One should 
however be aware of the fact that sometimes different choices are used,
especially for the exceptional groups.

Some values for the fundamental and adjoint representations are:
\begin{center}
\begin{tabular}{c|c|c|c|c|c|c|c|c|}
&SU(N)&SO(N)&SP(N)&$G_2$&$F_4$&$E_6$&$E_7$&$E_8$
\\ \hline
$C_A$    & $bN$        & $b(N-2)$   & $b(N+2)/2$ & 4$b$ & 9$b$ & 12$b$ & 18$b$ & 30$b$ \\
$I_4(A)$ & $2b^2N$     & $b^2(N-8)$ & $b^2(N+8)$ & 0  & 0  & 0   & 0   & 0   \\
$T_F$    & $b/2$       & $b$        & $b/2$      &  $b$ & 3$b$ & 3$b$  & 6$b$  &  -  \\
$N_F$    & $N$         & $N$        & $N$        &  7 & 26 & 27  &  56 &   - \\
$N_A$    & $N^2-1$     & $N(N-1)/2$ & $N(N+1)/2$ & 14 & 52 & 78  & 133 & 248 \\
\end{tabular}
\end{center}
For all groups we have $C_F = T_FN_A/N_F$. $I_4(F) = b^2$ for the classical 
groups and zero for the exceptional groups.
For other representations one would have to obtain values for the 
quantities $T_R$, $N_R$ and $I_4(R)$. By comparing the above values with the 
equations~\ref{rengroup} and ~\ref{eq:beta3} one may observe that the choice of 
a different value for b corresponds to a redefinition of the coupling constant.

\section*{Acknowledgements}
We are grateful to P.J. Nogueira and A.N. Schellekens for helpful and
stimulating discussions. The work of S.L. is
supported in part by the Russian Foundation
for Basic Research grant 96-01-01860.
The work of T.R. is supported by the US Department of Energy.
J.V. is grateful to the particle theory group of the University
of Michigan for its kind hospitality.


\begin{thebibliography}{99}
\bibitem{gvp}
D.J. Gross, F. Wilczek, Phys. Rev. Lett. 30 (1973) 1343; \\
H.D. Politzer, Phys. Rev. Lett. 30 (1973) 1346;\\
G. 't Hooft, report at the Marseille Conference on Yang-Mills Fields, 1972.
\bibitem{2l} W.E. Caswell, Phys. Rev. Lett. 33 (1974) 244; \\
D.R.T. Jones, Nucl. Phys. B 75 (1974) 531; \\
E.S. Egorian, O.V. Tarasov, Theor. Mat. Fiz. 41 (1979) 26.
\bibitem{tvz} O.V. Tarasov, A.A. Vladimirov, A.Yu. Zharkov.
Phys. Lett. B 93 (1980) 429.
\bibitem{lv} S.A. Larin, J.A.M. Vermaseren, Phys. Lett. B 303 (1993) 334.
\bibitem{h} G. 't Hooft, Nucl.Phys. B 61 (1973) 455.
\bibitem{hv} G. 't Hooft, M. Veltman, Nucl.Phys. B 44 (1972) 189;
\bibitem{abbott} L.F. Abbott, Nucl. Phys. B 185 (1981) 189.
\bibitem{collins} J.C. Collins, Nucl.Phys. B 80 (1974) 341.
\bibitem{bbdm} W.A. Bardeen, A.J. Buras, D.W. Duke, T. Muta, Phys. Rev. D 18
(1978) 3998.
\bibitem{vladimirov} A.A. Vladimirov, Theor.Mat.Fiz. 43 (1980) 210.
\bibitem{form} J.A.M. Vermaseren, Symbolic Manipulation with Form,
Computer Algebra Nederland, Amsterdam, 1991.
\bibitem{qgraph} P. Nogueira,  J. Comp. Phys.  105 (1993) 279.
\bibitem{qed}
S.G. Gorishny, A.L. Kataev, S.A. Larin, Phys.Lett.194B (1987) 429;\\
S.G. Gorishny, A.L. Kataev, S.A. Larin, 
L.R. Surguladze, Phys.Lett. B256 (1991) 81.
\bibitem{gracey} J.A. Gracey, Phys. Lett. B373 (1996) 178.
\bibitem{eks}M.A. Samuel, J. Ellis, M. Karliner,
Phys.Rev.Lett. 74 (1995) 4380;\\
J. Ellis, M. Karliner, M.A. Samuel, hep-ph/9612202 (1996).
\bibitem{patera}S. Okubo and J. Patera, J. Math Phys. 25 (1984) 219, ibid 24 (1983)
2722. 
\bibitem{cvitanovic} P.Cvitanovi\'c, Phys. Rev. D14 (1976) 1536.
\end{thebibliography}
\end{document}